\documentclass[prd,twocolumn,superscriptaddress,preprintnumbers,showpacs,nofootinbib,notitlepage,floatfix]{revtex4-1}
\usepackage{hyperref}
\usepackage{graphicx}
\usepackage{subfigure}
\usepackage{amsmath}
\usepackage{amssymb}
\usepackage{slashed}
\usepackage{amsfonts}
\usepackage{xcolor}
\usepackage{multirow}
\usepackage{array}
\usepackage{mwe}
\usepackage{mathrsfs}

\begin{document}

\title{Renormalon effect of quasi-PDF in gradient flow formalism}

\author{Jia-Lu Zhang}
\email{elpsycongr00@sjtu.edu.cn}
\affiliation{Tsung-Dao Lee Institute, Shanghai Jiao Tong University, Shanghai 201210, China}
\affiliation{State Key Laboratory of Dark Matter Physics, School of Physics and Astronomy, Shanghai Jiao Tong University,  Shanghai 200240, China}

\begin{abstract}
We investigate the behavior of renormalon ambiguities in quasi-parton distribution functions (quasi-PDFs) defined using gradient-flowed fields. Focusing on both ultraviolet (UV) and infrared (IR) renormalons, we perform an explicit calculation of bubble chain diagrams to analyze their dependence on the flow time. We find that the gradient flow eliminates UV renormalons in quasi-PDFs with zero external momentum, while IR renormalons persist but are significantly modified at finite flow time. The deformation of the renormalon structure induced by the gradient flow suggests an altered infrared behavior, with enhanced contributions from higher-twist effects even at small flow times.
\end{abstract}

\maketitle

\section{Introduction}

Lattice QCD provides a powerful non-perturbative framework for computing hadronic observables from first principles. However, two major challenges persist in practical lattice calculations. The first is the poor signal-to-noise ratio for certain correlation functions, especially those involving non-local operators or large external momenta, which are essential for accessing partonic structure. The second challenge lies in the renormalization of lattice observables. Standard continuum schemes such as $\overline{\mathrm{MS}}$ cannot be directly implemented on the lattice. Therefore, a lattice renormalization scheme should be employed for lattice observables. Among them, the regularization-independent momentum-subtraction scheme (RI/MOM) is often used~\cite{Constantinou:2017sej,Stewart:2017tvs}. However, the RI/MOM scheme is not gauge invariant and does not automatically eliminate linear divergences for non-local operators. Alternatively, self-renormalization and hybrid renormalization scheme have been developed~\cite{LatticePartonLPC:2021gpi,Ji:2020brr,Chou:2022drv}, but they depend heavily on fitting procedures and assumptions that may affect the reliability of the extracted renormalization factors. These issues limit the precision and reliability of observables extracted from lattice QCD.

A promising solution to both problems is provided by the gradient flow formalism~\cite{Narayanan:2006rf,Luscher:2010iy,Luscher:2011bx,Luscher:2013cpa}. This technique involves evolving the quantum fields along a fictitious flow-time direction, effectively smearing the fields over a finite physical scale. The flow suppresses ultraviolet (UV) fluctuations in a gauge-invariant manner, leading to significantly improved signal quality. 

These advantages have motivated the application of the gradient flow to the lattice computation of quasi-parton distribution functions (quasi-PDFs). In the Large Momentum Effective Theory (LaMET) framework, quasi-PDFs, which is defined as the Fourier transforms of equal-time correlators, can be matched onto standard PDFs through a factorization theorem~\cite{Ji:2013dva, Ji:2014gla, Izubuchi:2018srq}. Extensive work has been carried out to extract PDFs using LaMET and Lattice QCD, including the application of renormalization group equations (RGEs), higher-loop perturbative calculations, and various other techniques to improve the precision of the extraction~\cite{Xiong:2013bka,Lin:2014zya,Alexandrou:2015rja,Chen:2016utp,Alexandrou:2016jqi,Alexandrou:2018pbm,Chen:2018xof,Wang:2017qyg,Wang:2017eel,Lin:2018pvv,LatticeParton:2018gjr,Alexandrou:2018eet,Liu:2018hxv,Zhang:2018nsy,Izubuchi:2018srq,Izubuchi:2019lyk,Chen:2020arf,Chen:2020iqi,Chen:2020ody,Shugert:2020tgq,Chai:2020nxw,Lin:2020ssv,Fan:2020nzz,Gao:2021hxl,Gao:2021dbh,Gao:2022iex,Su:2022fiu,LatticeParton:2022xsd,Gao:2022uhg,Gao:2023ktu,Gao:2023lny,Chen:2024rgi,Holligan:2024umc,Holligan:2024wpv}.
It has been shown that flowed quasi-PDFs can be matched to quasi-PDFs renormalized in the $\overline{\mathrm{MS}}$ scheme~\cite{Brambilla:2023vwm, Monahan:2017hpu,Monahan:2016bvm}. Therefore, applying the gradient flow to quasi-PDFs can both enhance the signal-to-noise ratio and provide a systematic approach to renormalize lattice observables.

Nevertheless, a key challenge remains in controlling higher-twist effects. These power-suppressed contributions become significant at the finite hadron momenta accessible on the lattice and can obscure the extraction of leading-twist physics. A useful theoretical tool for analyzing such effects is the concept of renormalons~\cite{tHooft:1977xjm,Parisi:1978az,Parisi:1978bj,Beneke:1994sw,Beneke:1996gx,Beneke:1998ui,Beneke:2000kc,Han:2024cht,Braun:2024snf}. Renormalons arise from the divergent nature of perturbative expansions in QCD, where the coefficients grow factorially at high orders. This behavior leads to ambiguities in the Borel summation of the series, which reflect the presence of non-perturbative contributions. In particular, infrared (IR) renormalons are associated with long-distance dynamics and are expected to be canceled by the UV renormalons of higher-twist operators within the operator product expansion (OPE)~\cite{Beneke:1994sw}. Therefore, studying the renormalon structure of a given quantity provides insight into the pattern and magnitude of its power corrections.

Till now, plenty work has been done to study renormalon effect in effective theory or factorization theorem~\cite{Beneke:1995pq,Braun:2004bu,Gehrmann:2012sc,Scimemi:2016ffw,FerrarioRavasio:2020guj,Liu:2020rqi,Gracia:2021nut,Caola:2022vea,Makarov:2023ttq,Schindler:2023cww,Zhang:2023bxs,Schindler:2023cww,Mikhailov:2023gmo,Makarov:2023uet,Liu:2023onm,Han:2024cht,Zhang:2024wyq,Guo:2025obm}. The renormalon ambiguity of quasi-PDFs has been well-studied and has shed light on the nature of higher-twist contamination in LaMET-based factorization~\cite{Braun:2018brg,Braun:2024snf}. However, how renormalons manifest in flowed quasi-PDFs remains largely unexplored. Previous studies in the context of the gluon condensate have shown that UV renormalons are eliminated by the gradient flow, while IR renormalons remain unaffected~\cite{Suzuki:2018vfs,Beneke:2023wkq}. This behavior can be understood from the fact that the gradient flow suppresses UV fluctuations, thereby removing UV renormalons that arise from short-distance physics. In contrast, IR renormalons originate from long-distance dynamics and are therefore not influenced by the flow.

In this work, we investigate the renormalon ambiguity of flowed quasi-PDFs at zero external momentum within the large~$\beta_0$ approximation. Our analysis reveals that, provided the flow time~$t$ is finite, the IR renormalon structure of the quasi-PDFs is altered by the presence of the flow. Notably, the farther a Borel singularity is from the origin, the greater the deviation of its associated ambiguity from that in the unflowed case. This demonstrates that the gradient flow not only suppresses UV fluctuations but also modifies the IR renormalon structure at finite flow time.

The remainder of this paper is organized as follows. In Section II, we present the theoretical background essential to our analysis, including the gradient flow formalism as well as the definitions of quasi-PDFs and flowed quasi-PDFs. Section III reviews renormalon effects in quasi-PDFs and examines how these effects manifest within the gradient flow framework. Finally, Section IV offers a detailed study of renormalons in quasi-PDFs under gradient flow. Through explicit Borel analysis in simplified perturbative settings, we identify the structure of renormalon singularities and discuss their implications for extracting PDFs from flowed lattice data.

Our results provide new insights into the connection between perturbative and non-perturbative dynamics within the gradient flow framework and contribute to improving the reliability of parton structure determinations from lattice QCD.

\section{Theoretical Foundation}
\subsection{Gradient flow}
The gradient flow technique provides a  systematic framework to suppress UV fluctuations in both gauge and fermion fields~\cite{Narayanan:2006rf,Luscher:2010iy,Luscher:2011bx,Luscher:2009eq}. This powerful method has become indispensable in lattice QCD for scale setting and for computing local matrix elements and correlation functions with improved signal-to-noise ratios~\cite{Luscher:2013cpa,Luscher:2013vga,BMW:2012hcm,Sommer:2014mea,Suzuki:2013gza,Makino:2014taa,Harlander:2018zpi,FlavourLatticeAveragingGroupFLAG:2024oxs,Leino:2021vop,Mayer-Steudte:2022uih,Leino:2022kgj}. The method introduces a continuous evolution of quantum fields along a fictitious fifth dimension parameterized by the flow time $t$. Specifically, the original gauge field $A_\mu(x)$ and fermion field $\psi(x)$ transform into smoothed fields $B_\mu(t,x)$ and $\chi(t,x)$ through diffusion-like evolution equations~\cite{Luscher:2010iy,Luscher:2011bx,Luscher:2013cpa}.
\begin{equation}
\begin{aligned}
\label{flow}
\partial_t B_\mu &= D_\nu G_{\nu\mu} + \kappa D_\mu \partial_\nu B_\nu,  \\
\partial_t \chi &= D_\mu D_\mu \chi - \kappa (\partial_\mu B_\mu) \chi,\\
\partial_t \bar{\chi} &= \bar{\chi} \overleftarrow{D}_\mu \overleftarrow{D}_\mu + \kappa \bar{\chi} \partial_\mu B_\mu,    
\end{aligned}
\end{equation}
where \( D_\mu \) acts in the adjoint representation when applied to the gauge field and in the fundamental representation when applied to fermion fields. The parameter 
$\kappa$ is a gauge-fixing term that vanishes in physical observables. 
The field strength tensor is defined as
\begin{equation}
\begin{aligned}
G_{\mu\nu} &= \partial_\mu B_\nu - \partial_\nu B_\mu + [B_\mu, B_\nu],
\end{aligned}
\end{equation}
and the fields satisfy the following boundary conditions at $t=0$:
\begin{equation}
\begin{aligned}
B_\mu(x; t=0) &= g A_\mu(x), \\
\chi(x; t=0) &= \psi(x), \\
\bar{\chi}(x; t=0) &= \bar{\psi}(x).
\end{aligned}
\end{equation}

In the gradient flow formalism, operators evaluated at nonzero flow time \( t \neq 0 \) are known as ``flowed operators.'' As the flow time increases, Eq.~(\ref{flow}) drives the quantum fields toward classical configurations by effectively suppressing UV modes with wavelengths shorter than \( \sqrt{8t} \). This smoothing effect renders flowed composite operators UV finite, eliminating the need for additional renormalization.

The introduction of the flow time also defines a physical scale \( \sqrt{8t} \), which plays a key role in non-perturbative renormalization and scheme conversion. To faithfully extract physical observables, their characteristic scales must remain larger than this flow-induced cutoff.

The gradient flow can also be regarded as a continuous analogue of stout smearing~\cite{Luscher:2010iy,Luscher:2009eq,Nagatsuka:2023jos}. A significant advantage of the flow-based approach is that it enables perturbative computations of physical observables using modified Feynman rules~\cite{Lange:2021vqg}, whereas stout smearing is largely inaccessible to such analyses.

These features make the gradient flow a powerful tool in lattice QCD, enhancing signal-to-noise ratio and enabling reliable matching of lattice results to continuum schemes.

\subsection{Quasi-PDFs, Flowed Quasi-PDFs, and Matching}
Quasi-PDFs are constructed as the Fourier transformation of equal-time correlators
\begin{equation}
    \tilde{f}_q(x, P^z, \mu) = \int \frac{dz}{4\pi} \, e^{-i x P^z z}  \tilde{f}_q(z,P_z,\mu),
\end{equation}
with
\begin{equation}
\tilde{f}_q(z,P_z,\mu)=\langle P | \bar{\psi}_q(z) \gamma^z W[z, 0] \psi_q(0) | P \rangle.
\end{equation}
Here, $|P\rangle$ is a hadron state with momentum $P^\mu = (P^z,0,0,P^z)$, and $W[z, 0]$ is a spatial Wilson line along the $z$-direction.

In the framework of LaMET, the quasi-PDFs can be matched onto PDFs through a factorization theorem~\cite{Ji:2013dva, Ji:2014gla, Izubuchi:2018srq}
\begin{equation}
\begin{aligned}
\label{lamet_fact}
    \tilde{f}_q(x, P^z, \mu) = &\int_{-1}^{1} \frac{dy}{|y|} \, C\left(\frac{x}{y}, \frac{\mu}{P^z}\right) f_q(y, \mu) +\\ &C_2(x,\frac{\mu}{P^z})\otimes \mathcal{Q}_4(x,\mu)+...,
\end{aligned}
\end{equation}
where \( C(x/y, \mu/P^z) \) is a perturbatively calculable matching kernel that depends on the ratios \(x/y\) and \( \mu/P^z \), and  \( C_2(x, \mu/P^z) \otimes \mathcal{Q}_4(x,\mu)\) represents the twist-4 contribution in the factorization formula, giving rise to a correction of order \( \mathcal{O}(\Lambda_{\mathrm{QCD}}^2 / (P^z)^2) \). This approach allows for the first-principles extraction of PDFs by computing quasi-PDFs non-perturbatively with lattice QCD.

A further improvement involves the use of the gradient flow to the equal-time correlators $\tilde{f}_q(z,P_z,\mu)$ which defines the quasi-PDFs. It has been proven that the flowed equal-time correlators can be matched to the equal-time correlators in $\overline{\mathrm{MS}}$ at small flow time limit through a matching equation~\cite{Brambilla:2023vwm} 
\begin{equation}
\label{flow_fact}
\tilde f_q^{\mathrm{R}}(z,P^z,t) ={\cal C}(t, z, \mu) \,  e^{\delta mz}\tilde f_q^{\overline{\mathrm{MS}}}(z,P^z,\mu),
\end{equation}
where the matching kernel
\begin{equation}
\begin{aligned}
&\mathcal{C}_{\,t \to 0}(t,z, \mu) \\
&= 1 - \frac{\alpha_s}{4\pi} C_F \left[ 3 \log\left(2 \mu^2 t e^{\gamma_E}\right) + 2 + \log(432) \right]+\mathcal{O}(\alpha_s^2),
\end{aligned}
\end{equation}
is independent of $z$ at small flow time limit and
\begin{equation}
\delta m = -\frac{\alpha_s}{4\pi}C_F\frac{\sqrt{2\pi}}{\sqrt{t}} + \mathcal{O}(\alpha_s^2),
\end{equation}
represents the linear divergence arising from Wilson line self-energy corrections. In conventional regularization schemes, the linear divergence suffers from an intrinsic ambiguity associated with the UV renormalon. As we will demonstrate in the next section, the gradient flow formalism addresses this issue by using the flow time t as a physical regulator that removes the UV ambiguity.

\section{Renormalon Effects in Quasi-PDFs}
\begin{figure}
    \centering  
\includegraphics[width=1\linewidth]{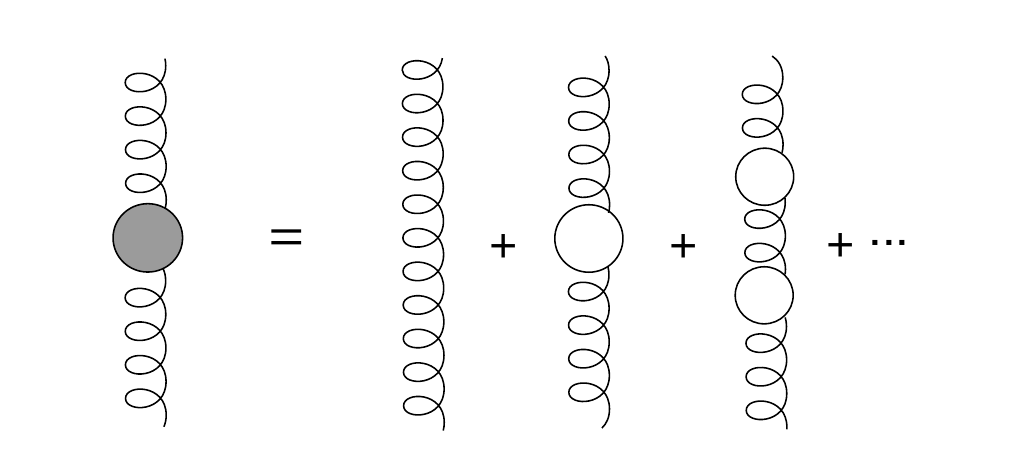}
\caption{The bubble chain diagrams correspond to changing a gluon propagator to gluon propagators dressed with fermion loops.}
\label{fig:bubble}
\end{figure}

The perturbative matching between quasi-PDFs and PDFs in the LaMET framework relies on the assumption that the difference between the two quantities is dominated by short-distance physics and can be systematically expanded in powers of $\alpha_s$ and $\Lambda_{\mathrm{QCD}}^2/(P^z)^2$. However, perturbative series in \( \alpha_s \) for QCD observables are known to be asymptotic and factorially divergent, due to the presence of IR and UV renormalons, which manifest as singularities in the Borel plane of the perturbative expansion~\cite{Parisi:1978bj,Beneke:1998ui}. These divergences signal intrinsic ambiguities in perturbation theory, associated with non-perturbative power corrections.
\begin{@twocolumnfalse}
\begin{center}
\begin{figure*}
    \centering  
\includegraphics[width=0.8\linewidth]{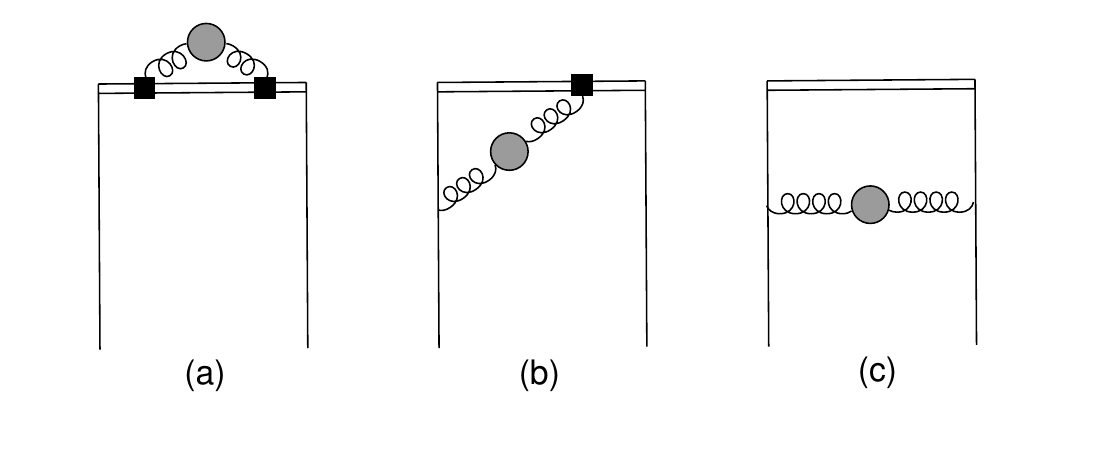}
\caption{Bubble diagrams contributing to the quasi-PDF. The black square denotes the flowed gluon field, the double line represents the Wilson line, and the gray bubble represents the bubble chain.}
\label{bubble_diagrams}
\end{figure*}
\end{center}
\end{@twocolumnfalse}
A standard method to assign meaning to asymptotic series is through Borel summation. To be explicit, consider an asymptotic expansion for a physical observable \( \phi(a_s) \), expressed in powers of \( a_s \equiv \frac{\alpha_s}{4\pi} \) as
\begin{equation}
\label{series}
    \phi(a_s) = \sum_{i=0}^{\infty} c_i \, a_s^{i+1},
\end{equation}
where the coefficients \( c_i \) are assumed to be known. The first step in Borel summation is to apply the Borel transform to the series, defined by
\begin{equation}
    \mathcal{B}[\phi](u) = \sum_{k=0}^{\infty} \frac{c_k}{k!} \left( \frac{u}{\beta_0} \right)^k,
\end{equation}
which maps the original asymptotic series into a new function of the Borel variable \( u \). The second step is to compute the Borel integral, given by
\begin{equation}
\label{Borel_Int}
    \Phi(a_s) = \frac{1}{\beta_0} \int_0^{\infty} \mathrm{d}u \, e^{-u / (\beta_0 a_s)} \, \mathcal{B}[\phi](u),
\end{equation}
where \( \beta_0 = 11 - \frac{2}{3}n_f \) is the one-loop coefficient of the QCD beta function. If the integral converges, the function \( \Phi(a_s) \) reproduces the same asymptotic expansion as the original series \( \phi(a_s) \), but yields a well-defined result that circumvents the divergence of the perturbative expansion. However, the integral in Eq.~(\ref{Borel_Int}) may be ill-defined due to singularities of \( \mathcal{B}[\phi](u) \) on the positive real axis. The ambiguity associated with choosing a contour to bypass these singularities must cancel against the ambiguity in the definition of higher-twist operators~\cite{Beneke:1998ui}.

In general, the factorial growth observed in matching kernels cannot be directly traced to individual Feynman diagrams. Nevertheless, it has been argued that the dominant contributions originate from a specific class of integrals known as ``bubble chain diagrams''~\cite{Gross:1974jv,Lautrup:1977hs,Zichichi:1977gu,Beneke:1998ui}. These diagrams are constructed by inserting multiple fermion loop corrections into a single gluon propagator within a one-loop diagram, a procedure referred to as the ``large \( \beta_0 \) approximation'', as illustrated in Fig.~\ref{fig:bubble}. This is followed by the replacement
\begin{align}
    -\frac{2}{3} n_f \to \beta_0 = \frac{11}{3}N_c - \frac{2}{3}n_f,
\end{align}
a step commonly known as ``naive non-Abelianization''~\cite{Broadhurst:1994se,Beneke:1994qe}.

The structure of renormalons in quasi-PDFs has been extensively studied using bubble chain diagrams~\cite{Braun:2018brg,Braun:2024snf}. These investigations reveal that the matching kernel contains IR and UV renormalons. The IR renormalons are associated with higher-twist corrections, suppressed by powers of \( \Lambda_{\mathrm{QCD}}^2 / (P^z)^2 \), while the UV renormalon primarily arise from the Wilson line self-energy and induce divergences linear in the Wilson line length.
 
The key question here is how renormalon effects are modified when gradient flow is applied to quasi-PDFs. This can be better understood through the factorization formula for the flowed quasi-PDFs:
\begin{equation}
\begin{aligned}
\label{comb_fact}
\tilde{f}_q(x, P^z, t) =& \int_{-1}^{1} \frac{dy}{|y|} \, C'\left(\frac{x}{y}, \frac{\mu}{P^z},t\right) f_q(y, \mu) + \\
&C_2'(x,\frac{\mu}{P^z},t)\otimes \mathcal{Q}_4(x,\mu)+...,
\end{aligned}
\end{equation} 
where $C'\left(\frac{x}{y}, \frac{\mu}{P^z},t\right) $ and $C_2'(x,\frac{\mu}{P^z},t)$ represent the matching kernels for the flowed quasi-PDFs.
It has been proposed~\cite{Suzuki:2018vfs,Beneke:2023wkq} that the gradient flow smooths UV fluctuations of quantum fields while leaving IR physics essentially unaffected. It leads to three consequences:
\begin{itemize}
    \item Comparing Eq.~(\ref{lamet_fact}) and Eq.~(\ref{comb_fact}), one finds that only the matching kernel is modified by the flow, while the non-perturbative matrix elements remain unchanged.
    \item As a result, the IR renormalons in the flowed and unflowed matching kernels must be identical, since they cancel against the UV renormalon ambiguities in the matrix elements.
    \item The UV renormalon ambiguity in the flowed matching kernel should be eliminated by the presence of a finite flow time.
\end{itemize}

The absence of the UV renormalon becomes evident in the bubble chain calculation: the flow time effectively introduces a UV cutoff in the Schwinger parameter integrals, thereby eliminating the associated UV divergence.

However, the validity of these statements should be confirmed through explicit numerical checks. In the following subsection, we investigate how the presence of renormalons is affected when quasi-PDFs are defined using gradient-flowed fields. For simplicity, we restrict our analysis to the case of quasi-PDFs evaluated at zero external momentum.

\section{Bubble chain results for flowed quasi-PDFs}

The bubble chain diagrams contributing to the zero-momentum flowed quasi-PDF are shown in Fig.~\ref{bubble_diagrams}. Although the Feynman rules are modified by the inclusion of flowed vertices in the gradient flow formalism, the analysis of bubble chain diagrams remains unaffected, as flowed vertices cannot be used to form closed fermion loops.

The explicit Borel-transformed result for Fig.~\ref{bubble_diagrams}~(a) reads:

\begin{equation}
\begin{split}
	&\mathcal{B}[\tilde{f}_a^R](w,t) = 
	C_F\, 2^{w-1} e^{5 w/3} t^{w-1} \mu^{2 w} \Gamma (-w) \\
    & \times \left[
    w z^2\, _2F_2\left(\tfrac{1}{2},1-w;\tfrac{3}{2},2; -\tfrac{z^2}{8 t}\right)
    - 4t \left(L_w\left(-\tfrac{z^2}{8 t}\right) - 1\right)
    \right],
\end{split}
\end{equation}
where \( L_w(z) \) denotes the Laguerre polynomial.
The first few residues of this Borel transform are
\begin{equation}
\begin{aligned}
&\operatorname{Res}_{w=1/2}\mathcal{B}[\tilde{f}_a^R] = 0, \\\quad
&\operatorname{Res}_{w=1}\mathcal{B}[\tilde{f}_a^R] = \tfrac{1}{2} e^{5/3} C_F \mu^2 z^2,\\ \quad
&\operatorname{Res}_{w=2}\mathcal{B}[\tilde{f}_a^R] = -\tfrac{1}{96} e^{10/3} C_F \mu^4 (96 t z^2 + z^4).   
\end{aligned}
\end{equation}
For comparison, the result in the \( \overline{\mathrm{MS}} \) scheme is
\begin{equation}
	\mathcal{B}[\tilde{f}_a^{\overline{\mathrm{MS}}}](w)= 
	\frac{C_F\, 2^{1-2 w} e^{5 w/3} \mu^{2 w} z^{2 w} \Gamma (-w)}{(2 w - 1)\, \Gamma(w + 1)},
\end{equation}
with corresponding residues
\begin{equation}
\begin{aligned}
&\operatorname{Res}_{w=1/2}\mathcal{B}[\tilde{f}_a^{\overline{\mathrm{MS}}}] = -2 e^{5/6} C_F \mu |z|, \quad \\
&\operatorname{Res}_{w=1}\mathcal{B}[\tilde{f}_a^{\overline{\mathrm{MS}}}] = \tfrac{1}{2} e^{5/3} C_F \mu^2 z^2, \quad\\
&\operatorname{Res}_{w=2}\mathcal{B}[\tilde{f}_a^{\overline{\mathrm{MS}}}] = -\tfrac{1}{96} e^{10/3} C_F \mu^4 z^4.
\end{aligned}
\end{equation}

Compared to the quasi-PDF in the \( \overline{\mathrm{MS}} \) scheme, the UV renormalon at \( w = \tfrac{1}{2} \) in Fig.~\ref{bubble_diagrams}~(a) is absent in the flowed quasi-PDF, while the IR renormalon at \( w = 2 \) acquires an additional term proportional to \( t \), which vanishes as \( t \to 0 \). These features are consistent with theoretical expectations: the introduction of a hard scale \( 1/\sqrt{t} \) implies that power corrections can appear not only in the form of \( z \Lambda_{\mathrm{QCD}} \), but also \( \sqrt{t} \Lambda_{\mathrm{QCD}} \). This can also be seen from the structure of the perturbative calculation, where such corrections naturally arise from the dimensionless ratio \( z^2 / 8t \), which commonly appears in gradient flow formulations.

At finite \( t \), deviations from the unflowed case become significant. This effect is even more pronounced in the results of Fig.~\ref{bubble_diagrams}~(b) and Fig.~\ref{bubble_diagrams}~(c).

The Borel-transformed result for Fig.~\ref{bubble_diagrams}~(b) is
\begin{equation}
\begin{aligned}
\mathcal{B}[\tilde{f}_b^R](w) &=C_F (-2^w) e^{5 w/3} t^w \mu^{2 w} \Gamma(-w) 
\\
&\times\left[_1F_1\left(-w;2; -\tfrac{z^2}{8 t}\right) - 1\right],  
\end{aligned}
\end{equation}
with the first three residues  
\begin{equation}
\begin{aligned}
\operatorname{Res}_{w=1} \mathcal{B}[\tilde{f}_b^R] &= -\tfrac{1}{8} e^{5/3} C_F \mu^2 z^2, \\
\operatorname{Res}_{w=2} \mathcal{B}[\tilde{f}_b^R] &= \tfrac{1}{192} e^{10/3} C_F \mu^4 (48 t z^2 + z^4), \\
\operatorname{Res}_{w=3} \mathcal{B}[\tilde{f}_b^R] &= -\tfrac{e^5 C_F \mu^6 }{9216}(2304 t^2 z^2 + 96 t z^4 + z^6).
\end{aligned}  
\end{equation}
The Borel-transformed results for Fig.~\ref{bubble_diagrams}~(c) is
\begin{equation}
\begin{aligned}
\mathcal{B}[\tilde{f}_c^R](w) &= \,
\frac{C_F\, e^{5 w/3} t^w \mu^{2 w} \Gamma(-w)}{2^{3-w}\Gamma(w+3)} \left[
3\, _1F_1\left(-w;3; -\tfrac{z^2}{8 t}\right)
\right.\\
& \left. + \,4 \left(4 \Gamma(w+3) - 1\right)\, _1F_1\left(-w;2; -\tfrac{z^2}{8 t}\right)
\right],
\end{aligned}
\end{equation}
with first three residues 
\begin{equation}
\begin{aligned}
\operatorname{Res}_{w=1} \mathcal{B}[\tilde{f}_c^R] &= 
\tfrac{e^{5/3} C_F \mu^2}{192}  (760 t + 47 z^2), \\
\operatorname{Res}_{w=2} \mathcal{B}[\tilde{f}_c^R] &= 
-\tfrac{e^{10/3} C_F \mu^4 }{73728}(294144 t^2 \\
&\qquad+ 36672 t z^2 + 763 z^4), \\
\operatorname{Res}_{w=3} \mathcal{B}[\tilde{f}_c^R] &= 
\tfrac{e^5 C_F \mu^6 }{22118400}(\,58951680 t^3\\
&+ 11047680 t^2 z^2 + 460200 t z^4 + 4793 z^6). 
\end{aligned}
\end{equation}
One observes that the coefficients associated with flowed operators can grow significantly as a function of the flow time. In the case of quasi-PDFs renormalized in the \( \overline{\mathrm{MS}} \) scheme, the renormalon ambiguity is suppressed when the corresponding Borel singularities are located farther from the origin. This behavior aligns with the general expectation that contributions from higher-dimensional operators are power-suppressed in a factorization framework.

However, in practical applications of gradient flow in lattice QCD, the flow time \( t \) must be fixed to a specific physical value before taking the continuum limit. This corresponds to the condition \( a^2 / t \to 0 \), where \( a \) is the lattice spacing. As a result, \( t \) is typically chosen to satisfy \( t > a^2 \). For example, a common choice is \( t \sim 0.03~\mathrm{fm}^2 \)~\cite{Brambilla:2023fsi}. In such cases, the large numerical coefficients can enhance the contributions of higher-dimensional operators, making them non-negligible.

These results admit several possible interpretations. First, the standard renormalon-based estimates of power corrections for lattice observables may not directly apply to flowed operators. This suggests that a more careful, possibly nonperturbative, study of power corrections within the gradient flow formalism is needed. Second, the modified IR structure induced by the flow could potentially enhance the so-called ``OPE nonconvergence'' phenomenon~\cite{Shifman:1994yf}. Finally, these observations raise the possibility that the large-\( \beta_0 \) approximation may break down or become less predictive in the presence of gradient flow.

It is important to emphasize that this calculation does not estimate power corrections arising from higher-twist light-cone operators, but rather from local higher-dimensional operators, due to the choice of zero external momentum~\cite{Braun:2018brg}. Nevertheless, it is natural to conjecture that this phenomenon extends beyond the zero-momentum matrix element and possibly beyond the specific case of quasi-PDFs.

\section{Conclusion and Outlook}

In this work, we investigated the renormalon structure of flowed quasi-PDFs within the framework of LaMET. The gradient flow, known for its ability to smooth UV fluctuations and define renormalized operators non-perturbatively, also provides a natural flow-time-dependent scheme for quasi-PDF calculations on the lattice. While it enhances signal-to-noise ratio and facilitates scheme conversion, its influence on non-perturbative power corrections has remained unclear.

To address this issue, we analyzed the Borel transform of flowed quasi-PDFs at zero external momentum and examined how renormalon ambiguities are modified by the presence of the flow time. Compared to the unflowed case, the flowed equal-time correlators are free of UV renormalons, and their IR renormalon structure is significantly altered, indicating a modified pattern of higher-twist corrections. Our analysis shows that the suppression of higher-twist effects, typically expected in the short-distance expansion, may be spoiled by the gradient flow if the flow time is kept finite—a common situation in practical lattice computations. This study can be extended to other key observables like quasi-GPDs and quasi-DAs within the LaMET framework or other factorization theorem~\cite{Han:2023hgy,Han:2024ucv,Deng:2023csv,LatticePartonLPC:2022eev,Han:2024cht,Guo:2025obm,Wang:2025uap,Han:2025odf,LatticeParton:2024vck,Han:2023xbl,Hu:2023bba,LatticeParton:2023xdl,LatticePartonCollaborationLPC:2022myp,Deng:2022gzi,LatticeParton:2022zqc,Hua:2020gnw,LatticeParton:2020uhz,Wang:2019msf,Wang:2019tgg,LatticeParton:2018gjr,Zhang:2023bxs,Wang:2024wwa,Wang:2025uap,LatticeParton:2024zko,Han:2024fkr}, enabling enhanced error control and improved systematic accuracy in their extraction.

It is important to note that our calculation focuses on the zero-momentum limit, thereby probing the power corrections relevant to the short-distance OPE, rather than those arising in the light-cone OPE relevant for quasi-PDFs at large momentum. In future work, we plan to extend our analysis to the quasi-PDF with large external momentum.

\appendix
\section{Explicit Calculation for Bubble chain diagrams}

In this appendix, we present the bare results for the three bubble chain diagrams shown in Fig.~\ref{bubble_diagrams}, including a detailed calculation for Fig.~\ref{bubble_diagrams}~(a).
\begin{widetext}
The result for FIG.~\ref{bubble_diagrams}~(b) reads
\begin{equation}
\begin{aligned}
    \tilde{f}_b^R=-\frac{C_F f^n \beta _0^n 2^{(n+1) \epsilon } e^{\gamma  (n+1) \epsilon } \epsilon ^{-n} a_s^{n+1} t^{(n+1) \epsilon } \mu ^{2 (n+1) \epsilon }
   \Gamma (-((n+1) \epsilon )) \left(\, _1F_1\left(-((n+1) \epsilon );2-\epsilon ;-\frac{v^2 z^2}{8 t}\right)-1\right)}{\Gamma (2-\epsilon )}.
\end{aligned}
\end{equation}
The result for FIG.~\ref{bubble_diagrams}~(c) reads
\begin{equation}
\begin{aligned}
\tilde{f}_c^R&=\frac{C_F (\epsilon -1) f^n \beta_0^n 2^{n \epsilon +\epsilon -2} e^{\gamma  (n+1) \epsilon } \epsilon ^{-n} a_s^{n+1} t^{(n+1) \epsilon} \mu ^{2 (n+1) \epsilon } \Gamma (-((n+1) \epsilon ))}{\Gamma (n \epsilon +3)} \\
&\times \left((2 \epsilon -3) \, _1\tilde{F}_1\left(-((n+1) \epsilon );3-\epsilon ;-\frac{z^2}{8 t}\right)\right. 
\left.+(2-8 \Gamma (n \epsilon +3)) \, _1\tilde{F}_1\left(-((n+1) \epsilon );2-\epsilon ;-\frac{z^2}{8 t}\right)\right).
\end{aligned}
\end{equation}
The calculation for Fig.~\ref{bubble_diagrams}~(a) begins with the bare bubble chain expression 
\begin{equation}
\begin{aligned}
\tilde{f}^R_{a}&=(ig)^2\int_0^z ds_1\int_0^{s_1}ds_2 \langle P|\bar{\psi}_i(zv,t)\Gamma  v\cdot {A^i}_{j}(v_1s,t) v\cdot {A^j}_{i}(v_2s,t)\psi^i(0,t)|p\rangle \\
    &=-\int d^4k\int_0^z ds_1\int_0^{s_1}ds_2\int_0^\infty d\sigma\frac{C_F g^2 2^{-d-2 \epsilon } \pi ^{-d-\epsilon } f^n \beta _0^n \epsilon
   ^{-n} a_s^n \mu ^{2 n \epsilon +2 \epsilon } \sigma ^{n \epsilon } \exp \left(-k^2
   \sigma -2 k^2 t+i k v (s_2-s_1)+\gamma  n \epsilon +\gamma  \epsilon
   \right)}{\Gamma (n \epsilon +1)}\\
   &=-\int_0^z ds_1\int_0^{s_1}ds_2\int_0^\infty d\sigma\frac{C_F 2^{-d} \pi ^{-d/2} g^2 f^n \beta _0^n \epsilon ^{-n} a_s^n (\sigma +2
   t)^{-d/2} \mu ^{2 n \epsilon } \sigma ^{n \epsilon } \exp \left(\gamma  (n+1)
   \epsilon -\frac{(s_1-s_2)^2}{4 (\sigma +2 t)}\right)}{\Gamma (n
   \epsilon +1)}\\
   &=-\frac{C_F \, f^n \beta_0^n \, 2^{n\epsilon + \epsilon -1} \, e^{\gamma (n+1)\epsilon} \, \epsilon^{-n-1} \, a_s^{n+1} \, t^{n\epsilon + \epsilon -1} \, \mu^{2(n+1)\epsilon} \, \Gamma(1-(n+1)\epsilon)}{(n+1) \Gamma(2-\epsilon)} \notag \\
& \quad \times \left( (n+1) z^2 \epsilon \, {}_2F_2\left(\tfrac{1}{2}, -n\epsilon - \epsilon + 1; \tfrac{3}{2}, 2 - \epsilon; -\tfrac{z^2}{8t}\right) + 4t(\epsilon - 1) \left({}_1F_1\left(-((n+1)\epsilon); 1 - \epsilon; -\tfrac{z^2}{8t}\right) - 1\right) \right).
\end{aligned}
\end{equation}
We then renormalize the bare bubble chain diagram. A bare diagram containing \( n \) fermion loops is replaced by a sum over diagrams in which some of the fermion loops are substituted with their corresponding counterterms. Terms proportional to \( \epsilon^{-n} \) with \( n > 0 \) are discarded, as they are canceled by renormalization. This substitution leads to factorial growth in the perturbative expansion, since each distinct permutation of fermion loops and counterterms contributes to the overall \( n! \) factor

\begin{equation}
\begin{aligned}
\tilde{f}_a^R/a_s^{n+1}=&\sum_{k=0}^n -C(n,k) \, \frac{C_F \, (-1)^k \,  \, f^{n-k} \left( \frac{\beta_0}{\epsilon} \right)^k \beta_0^{n-k} \, 2^{-k\epsilon + n\epsilon + \epsilon -1} \, e^{\gamma \epsilon (-k + n + 1)} \, \epsilon^{k - n - 1}}{(-k + n + 1) \Gamma(2 - \epsilon)}\\
& \times t^{-k\epsilon + n\epsilon + \epsilon - 1} \, \mu^{2\epsilon(-k + n + 1)} \, \Gamma((k - n - 1)\epsilon + 1) \left( z^2 \epsilon (-k + n + 1) \, {}_2F_2\left(\tfrac{1}{2}, k\epsilon - n\epsilon - \epsilon + 1; \tfrac{3}{2}, 2 - \epsilon; -\tfrac{z^2}{8t} \right) \right. \notag \\
&\left. + \, 4t(\epsilon - 1) \left({}_1F_1\left((k - n - 1)\epsilon; 1 - \epsilon; -\tfrac{z^2}{8t} \right) - 1 \right) \right)\\
=&\sum_{k=0}^n C(n,k) (-1)^k \left( \frac{\beta_0}{\epsilon} \right)^n 
\frac{C_F \, f^{n-k} \, 2^{-k\epsilon + n\epsilon + \epsilon - 1} \, e^{\gamma \epsilon (-k + n + 1)} \, t^{-k\epsilon + n\epsilon + \epsilon - 1} \, \mu^{2\epsilon (-k + n + 1)}}{\epsilon (k - n - 1) \Gamma(2 - \epsilon)}\\
& \times \Gamma((k - n - 1)\epsilon + 1) 
\left( z^2 \epsilon (-k + n + 1) \, {}_2F_2\left( \tfrac{1}{2}, k\epsilon - n\epsilon - \epsilon + 1; \tfrac{3}{2}, 2 - \epsilon; -\tfrac{z^2}{8t} \right) \right. \\
&  \left. + \, 4t(\epsilon - 1) \left( {}_1F_1\left( (k - n - 1)\epsilon; 1 - \epsilon; -\tfrac{z^2}{8t} \right) - 1 \right) \right).
\end{aligned}
\end{equation}
We then change the varible $s=(n-k+1)\epsilon$,
\begin{equation}
\begin{aligned}
\tilde{f}_a^R/a_s^{n+1}&= -\sum_{k=0}^n C(n,k)\left(\frac{\beta_0}{\epsilon}\right)^n (-1)^k 
\frac{C_F 2^{s-1} e^{\gamma s} \mu^{2s} t^{s-1} \Gamma(1-s) f^{\frac{s}{\epsilon}-1}}{s\, \Gamma(2-\epsilon)} \\
&\quad \times \left[ s z^2 \, {}_2F_2\left(\frac{1}{2},1-s;\frac{3}{2},2-\epsilon; -\frac{z^2}{8t}\right) 
+ 4t(\epsilon -1) \left( {}_1F_1\left(-s;1-\epsilon; -\frac{z^2}{8t}\right) - 1 \right) \right].
\end{aligned}
\end{equation}
To simplify the expression, one can set
\begin{equation}
\begin{aligned}
    G(\epsilon,s)&=-\frac{C_F 2^{s-1} e^{\gamma  s} \mu ^{2 s} t^{s-1} \Gamma (1-s) f^{\frac{s}{\epsilon }-1} \left(s z^2 \,
_2F_2\left(\frac{1}{2},1-s;\frac{3}{2},2-\epsilon ;-\frac{z^2}{8 t}\right)+4 t (\epsilon -1) \left(\, _1F_1\left(-s;1-\epsilon ;-\frac{z^2}{8
t}\right)-1\right)\right)}{s \Gamma (2-\epsilon )},
\end{aligned}
\end{equation}
\end{widetext}
together with it's Laurent expansion
\begin{equation}
\tilde{f}_a^R/a_s^{n+1}= \beta^n
   \left(\frac{1}{\epsilon}\right)^n\sum_i\sum_{k=0}^nC(n,k)(-1)^k(g_i^{[n]}s^n)\epsilon^i.
 \end{equation}
Then we arrives at 
\begin{equation}
	\tilde{f}_a^R/a_s^{n+1}= \beta^n
   \left(\frac{1}{\epsilon
   }\right)^n\sum_{k=0}^nC(n,k)(-1)^kG(\epsilon,s).
 \end{equation}

Using the formulas related to the second Stirling  numbers:
\begin{align}
 & 
\begin{aligned}
\sum_{k=0}^n\frac{(-1)^k\binom{n}{k}}{(n-k+1)^2}=\frac{(-1)^n}{n+1}H_{n+1},
\end{aligned} \\
 & 
\begin{aligned}
\sum_{k=0}^n\frac{(-1)^k\binom{n}{k}}{(n-k+1)}=\frac{(-1)^n}{n+1},
\end{aligned} \\
 & \sum_{k=0}^n(-1)^k\binom{n}{k}(n-k+x)^j=0,0\leq j\leq n-1,\forall x, \\
 & \sum_{k=0}^n(-1)^k\binom{n}{k}(n-k+x)^n=n!,\forall x, & 
\end{align}
and do the Borel transformation, one finally gets
\begin{equation}
\begin{aligned}
&\mathcal{B}[\tilde{f}_a^R](w)= \sum_{n=0}^{\infty}g_0^{[n]} w^n=G(0, w)\\
&= C_F\, 2^{w-1} e^{5 w/3} t^{w-1} \mu^{2 w} \Gamma (-w) \\
    & \times \left[
    w z^2\, _2F_2\left(\tfrac{1}{2},1-w;\tfrac{3}{2},2; -\tfrac{z^2}{8 t}\right)
    - 4t \left(L_w\left(-\tfrac{z^2}{8 t}\right) - 1\right)
    \right].
\end{aligned}
 \end{equation}

\section*{Acknowledgement}
The author gratefully acknowledges Prof.~Wei Wang and Dr.~Chao Han for valuable discussions and insightful comments on the results. Thanks are also due to Prof.~Nora Brambilla and Prof.~Xiang-Peng Wang for helpful discussions on the gradient flow formalism. This work is supported by the T. D. Lee Scholarship.

\end{document}